\documentclass[a4paper,twoside]{article}
\newcounter{the_style}
\input{main.macros}
\author{Jochen Br\"{o}cker\thanks{%
Max--Planck--Institut f\"{u}r Physik komplexer Systeme, N\"{o}thnitzer Strasse~34, 01187 Dresden, Germany, email: \texttt{broecker@pks.mpg.de}}%
}
\title{Generating Probabilities From Numerical Weather Forecasts by Logistic Regression}
\begin{document}
\maketitle
\begin{abstract}
Logistic models are studied as a tool to convert output from numerical
weather forecasting systems (deterministic and ensemble) into probability
forecasts for binary events.
A logistic model obtains by putting the logarithmic odds ratio equal to a linear combination of the inputs. 
As any statistical model, logistic models will suffer from over-fitting if the number of inputs is comparable to the number of forecast instances. 
Computational approaches to avoid over-fitting by regularisation are discussed, and efficient approaches for model assessment and selection are presented.
A logit version of the so called lasso, which is originally a linear tool, is discussed.
In lasso models, less important inputs are identified and discarded, thereby providing an efficient and automatic model reduction procedure. 
For this reason, lasso models are particularly appealing for diagnostic purposes.
\end{abstract}

\section{Introduction}
\label{sec:introduction}
Providing forecasts of future events in terms of probabilities has a long and successful history in the environmental sciences. 
The inherently unstable dynamics of the atmosphere in conjunction with incomplete information on its current state prohibit unequivocal forecasts.
Probabilities allow to quantify uncertainty (or the absence thereof, i.e.\ information) in a well defined and consistent manner. 
So called ``subjective probability forecasts'', compiled by experienced meteorologists, were issued by several meteorological offices from the 1950's onwards.
These forecasts were based on synoptic weather data collected from a large number of stations.
On a scientific (non--operational) level, probabilistic weather forecasts were discussed much earlier, either based on synoptic information as well as local station data~\citep[see e.g.][provides an excellent account on the history of probability forecasting of weather, along with many references]{besson05,murphy98}.
There is evidence that subjective probabilistic weather forecasts were compiled from non--synoptic information as early as the 1790's~\citep[][]{murphy98}.
Desirable properties of probabilistic forecasts as well as methods to quantify their success have been investigated in various papers, see for example~\citet{myrbach13,brier50,good52,winkler68,murphy77,murphy87,murphy93,murphy96}, a list which is by no means complete.
The advent of the electronic computer opened the possibility to calculate ``numerical subjective probabilities'', that is, to calculate probabilities from information data using tuned algorithms.
More specifically, an automated statistical learning procedure can be employed to find a relationship (also called {\em model}) between the information data (also called {\em covariates}, {\em features}, or {\em inputs}) and the statistical properties of the variable to be forecast (also called {\em target}, or {\em verification}).
Possible inputs might for example be provided by output from a numerical weather prediction system, in which case the problem is also referred to as ensemble calibration or probabilistic down-scaling.
A model (or more specifically model {\em class}) which has gained some attention in the meteorological community is the logistic model, see for example~\citet{tippet06,wilks06,hamill04} and also references therein for various alternatives.
Logistic models, often also referred to as logistic regression, will be the subject of this paper.
We will exclusively be concerned with dichotomic problems, that is, we are only interested in forecasting whether a certain event happens or not.
In this case, the logistic model obtains by taking the logarithmic odds ratio $\log (\frac{\rho}{1 - \rho})$ of the forecast probability $\rho$ of the event as a linear function of the inputs.
In other words,
\beq{equ:10}
\rho = \frac{\exp(x\beta^t)}{1 + \exp(x\beta^t)},
\eeq
where $x$ are the inputs and $\beta$ some coefficients.
The maximum likelihood principle provides a convenient way to find the coefficients, but alternatives will be considered in this paper, too.
In any case, the coefficients are found by optimising the performance over some training data.
As with other regression models though, this approach runs into problems if the number of inputs is of the same order of magnitude as the number of instances in the training data, in which case the inputs are typically also highly correlated.
This is a common situation in weather forecasting, owing to the large number of available forecast sources. 
One way to avoid over-fitting in this situation is to manually restrict the number of inputs to the few that seem to be most relevant.
This was carried out for example by \citet{besson05}, but often such a study would require to assess an astronomically large amount of different combinations.
Another way is to apply {\em regularisation}, which means to reduce the effective degrees of freedom of the model.
Efficient regularisation techniques exist for linear models.
Owing to the great similarity between linear and logistic models, these techniques can be modified and applied to logistic models, as will be demonstrated in this paper. 
Logistic models will be defined in Section~\ref{sec:problem_statement}.
Section~\ref{sec:logit_regular} discusses how to regularise logistic models along with further computational and implementational aspects.
Sections~\ref{sec:problem_statement} and~\ref{sec:logit_regular} therefore form the core of the paper.
Before getting there, some notation and concepts need to be introduced (Section~\ref{sec:problem_statement}).
Numerical studies employing precipitation forecasts are presented in Section~\ref{sec:numerical_studies}.
Apart from the predictive power of logistic models, they also permit to investigate the relative importance of the inputs.
The material of this last section thus also demonstrates the capabilities of logistic models as a diagnostic tool.
\section{Problem Statement and Concepts}
\label{sec:problem_statement}
The primary aim of this section is to settle notational conventions and introduce some concepts.
The general setup we have in mind is as follows.
As in the introduction, let the target $Y$ be a random variable taking the values $0$ and $1$ only, with $Y = 1$ indicating that the event under concern happened and $Y = 0$ otherwise. 
The inputs $X$ are random variables too, taking values in $\R^d$.
By $x$, we will denote any generic point in $\R^d$, while $y$ is always either zero or one. 
The underlying probability measure will be denoted by $\P$.
The probabilistic relationship between $X$ and $Y$ is described through the following objects.
Let
\begin{equation}
\label{equ:likelihoods}
\begin{split}
f_0(x) & := \P(X \in x + \dd x | Y = 0), \\
f_1(x) & := \P(X \in x + \dd x | Y = 1),
\end{split}
\end{equation}
that is, $f_0$ and $f_1$, respectively, are the densities of $X$ given $Y = 0$ and $Y = 1$, respectively.
By
\begin{equation}
\label{equ:condprob}
\pi(x) := \P(Y = 1| X = x)
\end{equation}
we denote the conditional probability of the event ``$Y = 1$'' given $X$, and
\begin{equation}
\label{equ:baserate}
\bar{\pi} := \P(Y = 1)
\end{equation}
denotes the {\em base rate} or  {\em grand probability} of the event ``$Y = 1$''.
Finally,
\begin{equation}
\label{equ:featuredistr}
f(x) := \P(X \in x + \dd x)
\end{equation}
denotes the unconditional density of the inputs $X$.
The Bayes rule entails various relations between these objects, for example
$
f(x) = f_1(x) \bar{\pi} + f_0(x) (1 - \bar{\pi}).
$
A {\em model} is a function $\rho: \R^d \to [0, 1]$ so that $\rho(X)$ is the forecast probability of the event ``Y = 1''.
Generally speaking, the problem discussed in this paper is to find ``good'' models, where it remains to be defined what ``good'' means.
Intuitively, it is clear that $\rho(x) := \pi(x)$ is a good model. 
Unfortunately, $\pi(x)$ is not an empirically measurable quantity, and therefore ``interpolating'' or ``fitting'' $\pi(x)$ is not a possible approach to determine $\rho(x)$.
A general operational criterion to fit (deterministic or probabilistic) relationships between measured quantities is to optimise the estimated performance, according to suitable criteria of performance.
Such criteria are subject of the next subsection.

\subsection{Scoring Rules\label{subsec:scoring_rules}}
A {\em scoring rule}~\citep{good52,kelly56,brown70,savage71} is a function $S(p, y)$ where $p \in [0, 1]$ and $y$ is either zero or one. If $\rho(X)$ is the forecast probability and $Y$ is the corresponding target, then $S(\rho(X), Y)$ quantifies how well $\rho(X)$ succeeded in forecasting $Y$.
Two important examples are the Ignorance score~\citep{good52}, given by the scoring rule
\begin{equation}
\label{equ:score30}
S(p, y) := -\log(p) \cdot y - \log(1 - p) \cdot (1 - y),
\end{equation}
and the Brier score~\citep{brier50}, given by the scoring rule
\begin{equation}
\label{equ:score40}
S(p, y) := (y - p)^2 = (1 - p)^2 \cdot y + p^2 \cdot (1 - y). 
\end{equation}
These definitions imply the convention that a smaller score indicates a better forecast.
A score or scoring rule quantifies the success of individual forecast instances by comparing the random variables $\rho(X)$ and $Y$ point-wise.
The general quality of a forecasting system is commonly measured by the mathematical expectation $\Ex \left[ S(\rho(X), Y) \right]$ of the score, which can be estimated by the empirical mean
\begin{equation}
\label{equ:score70}
\Ex \left[ S(\rho(X), Y) \right]
\cong \frac{1}{N} \sum_{i = 1}^N S(\rho(x_i), y_i)
\end{equation}
over a sufficiently large set $\{ (x_i, y_i); i = 1\ldots N \}$ of input--target pairs.
Reassuringly, for the two mentioned scoring rules the score becomes better
(i.~e.~decreases) with increasing $\rho$ if the event happened, while if it does
not, the score becomes worse (i.e.\ {\em in}creases) with increasing $\rho$. 
Furthermore, both scores are {\em proper}. 
To define this notion, consider the {\em scoring function}
\begin{equation}
\label{equ:score10}
s(q, p) := S(q, 1) \cdot p + S(q, 0) \cdot (1 - p)
\end{equation}
where $q, p$ are two arbitrary probabilities, that is, numbers in the unit interval.
The scoring function is the mathematical expectation of the score in a situation where the forecast is $q$ but in fact $p$ is the true probability of the event ``$Y = 1$''.
A score is {\em strictly proper}~\citep{brown70,broecker06-3} if the {\em divergence function} (or loss function)
\begin{equation}
\label{equ:score20}
d(q, p) := s(q, p) - s(p, p)
\end{equation}
is positive definite, that is, never negative and zero only if $p = q$.
The divergence function of the Brier score for example is $d(q, p) := (q - p)^2$, demonstrating that this score is strictly proper.
The Ignorance is proper as well, since \eqref{equ:score20} is just the Kullback--Leibler--divergence, which is well known to be positive definite.
The mathematical expectation of a strictly proper score allows for a very interesting decomposition \citep[see][for a proof]{broecker07-3}.
For any strictly proper scoring rule, define the {\em entropy} $e(p) := s(p, p)$.
Furthermore let $\pi_{\rho}(r) := \P(Y = 1 | \rho(X) = r) $ be the conditional probability of $Y = 1$ given that $\rho(X) = r$. 
This quantity is a function of $\rho$, but it is a fully calibrated probability forecast.
With these definitions, it can be shown that
\begin{equation}
\label{equ:score80}
\Ex S(\rho, Y) = e(\bar{\pi}) - \Ex d(\bar{\pi}, \pi) 
+ \Ex d(\pi_{\rho}, \pi)
+ \Ex d(\rho, \pi_{\rho}).
\end{equation}
These terms can be interpreted as follows: The entropy $e(\bar{\pi})$ is the ability of the base rate $\bar{\pi}$ to forecast draws from itself, and hence quantifies the fundamental uncertainty inherent in $Y$.
The term $\Ex d(\bar{\pi}, \pi)$ is positive definite and quantifies the average divergence of $\pi$ from its mean. It can hence be considered a generalised variance of $\pi$. If the Brier score is used, this term is in fact the ordinary variance of $\pi$.
The term $\Ex d(\pi_{\rho}, \pi)$ is also positive definite and quantifies how much information is lost when going over from $X$ to $\rho(X)$.
The term $\Ex d(\rho, \pi_{\rho})$ is again positive definite and quantifies the imperfect calibration of $\rho$.
The reader might want to convince himself that if the Brier score is used and furthermore $X = \rho(X)$ (i.e.\ the inputs already comprise a probability forecast), then relation~\eqref{equ:score80} agrees with the well known decomposition of the Brier score.
In particular, if $\rho(x) = \pi(x)$, then also $\pi_{\rho} = \pi(x)$ and hence the third and fourth term in Equation~\eqref{equ:score80} vanish. 
We can conclude that the forecast $\pi(X)$ in fact yields an optimum expected score among all models which can be written as a function of $X$. 
To achieve yet better scores, more information about $Y$ is needed than what is contained in $X$.

\subsection{Logistic Regression}
Comprehensive discussions of Logistic regression can be found in~\citet{mccullagh89,hastie01}. As mentioned in the Introduction, logistic regression assumes a model of the form $ \rho(x) = g(x\beta^t)$, where~$\beta$ are the coefficients and
\beq{equ:5.30}
g(z) := \frac{\exp(z)}{ 1 + \exp(z)}
\eeq
the so--called {\em link function}. 
The quantity $g^{-1}(z) = \log(z / (1 - z))$ is referred to as the {\em log--odds--ratio}.
We will call $ \eta := x\beta^t$ the linear response.
From now on and throughout the paper, we will assume that the inputs $x$ carry an entry $1$ in the first position and that $\beta = (\beta_0 \ldots \beta_{d})$ where $\beta_0$ is referred to as the {\em intercept}.
Since $g^{-1}(\rho) = \eta$, in logistic models, the log--odds--ratio equals the linear response.
The coefficients~$\beta$ are determined by minimising the empirical score.
Locally around the optimum, this minimisation turns out to be equivalent to weighted linear regression, as will be seen in the next subsection.
Thereby, logistic models inherit various useful properties from linear models, as long as strictly proper scores are used in the empirical score.
This fact will be exploited in the next section.
\section{Computational Topics and Regularisation of Logistic Models}
\label{sec:logit_regular}
Consider the empirical score of a logistic model
\beq{equ:6.10}
R(\beta) := \frac{1}{N}\sum_{k = 1}^N S(g(x_k \beta^t), y_k)
\eeq
where as before $S$ is a scoring rule, $g$ is the link function, and $\{(x_k, y_k), k = 1 \ldots N\}$ is a set of input--target pairs, henceforth called {\em training set}.
We let $\hat{R}$ denote the minimum of $R$ with respect to $\beta$, and $\hat{\beta}$ a corresponding stationary point.
To find a stationary point of $R$, the Newton--Raphson algorithm can be used. The update step for this iterative algorithm can be written as
\beq{equ:6.20}
\beta_{new}^t = \beta^t - \left( \mtr{X}^t \mtr{W} \mtr{X} \right)^{-1} \* \mtr{X}^t \mtr{d}
\eeq
with the abbreviations
\begin{eqnarray}
\label{equ:6.30}
\mtr{X}_{kl} & := & x_k^{(l)}\\
\label{equ:6.40}
\mtr{d}_{k} & := &  \frac{\partial}{\partial \eta} S(g(\eta_k), y_k)\\
\label{equ:6.45}
\mtr{w}_{k} & := &  \frac{\partial^2}{\partial \eta^2} S(g(\eta_k), y_k)\\
\label{equ:6.50}
\mtr{W}_{kl} & := & \delta_{kl} \mtr{w}_{k},
\end{eqnarray}
where $\eta_k := x_k \beta^t$ is the log--odds--ratio for sample $k$.
In the case of the Ignorance, it is easy to see that
\beq{equ:6.55}
\mtr{d}_{k} =  g(\eta_k) - y_k , \qquad
\mtr{w}_{k} = g(\eta_k)(1 - g(\eta_k))
\eeq
Equation~\eqref{equ:6.20} is in fact very similar to weighted least squares regression with linear models~\citep{hastie01}.
\subsection{$L_2$--Type Regularisation}
\label{subsec:l_2-type-regularisation}
Whichever score or minimisation algorithm we employ, the coefficients $\hat{\beta}$ so determined will show poor out--of--sample performance if the number of degrees of freedom is of the same order of magnitude as the number of instances in the training set.
Small changes in the training set will entail large changes in the coefficients, or in other words, the coefficients will exhibit large variance. 
Apart from poor performance, the results become difficult to interprete~\citep{hastie01}.
To give a heuristic argument as to why this happens, suppose we want to fit a linear model to real valued data, and there are two highly correlated inputs. 
Any large value for $\beta_1$ (the coefficient corresponding to the first input) can be compensated by a large value (with opposing sign) for $\beta_2$ (the coefficient corresponding to the second input).
The algorithm will use this freedom to ``model'' the random fluctuations in the inputs.
To avoid this behaviour, the degrees of freedom of the model have to be reduced (or, what amounts to the same, the range of variation of $\beta$), preferably in an adaptive manner.
A straight forward approach is to search for a minimum of $R$ only among all $\beta$ with $|\beta|^2 \leq \mu$ with a {\em regularisation parameter} $\mu$.
Here $|\beta|^2 = \sum_{k = 1}^{d} \beta_k^2$.
The reader is reminded of the convention that $\beta$ has $d+1$ entries in total, but we decided not to put any constraint on the intercept $\beta_0$.
Furthermore, we assume that all inputs are centred and scaled so that they have mean zero and unit variance.
To see why regularisation has the desired effect, and to obtain criteria for choosing an appropriate $\mu$, the problem has to be brought into another form.
For $\hat{\beta}$ to be a stationary point of $R$ under the constraint $|\beta|^2 \leq \mu$, it is necessary that there is a $\hat{\lambda}$ so that the Lagrangian
\[
L(\beta, \lambda) := R(\beta) + \lambda ( |\beta|^2 - \mu)
\]
has a saddle at $(\hat{\beta}, \hat{\lambda})$, which obviously entails that $\hat{\beta}$ is a stationary point for 
\beq{equ:6.60}
R_{\hat{\lambda}}(\beta) := R(\beta) + \hat{\lambda} |\beta|^2.
\eeq
We arrive at the conclusion 
\begin{quote}
If $\hat{\beta}$ maximises $R(\beta)$ under the constraint $|\beta|^2 \leq \mu$ and if $\hat{\lambda}$ is the corresponding Lagrange multiplier, then $\hat{\beta}$ is a stationary point of $R_{\hat{\lambda}}(\beta)$. Conversely, if we fix a $\hat{\lambda} > 0$ and let $\hat{\beta}$ be a stationary point of $R_{\hat{\lambda}}(\beta)$, then $\hat{\beta}$ maximises $R(\beta)$ under the constraint $|\beta|^2 \leq \mu := |\hat{\beta}|$, and the corresponding Lagrange multiplier is $\hat{\lambda}$.
\end{quote}
It is probably more intuitive to optimise $R(\beta)$ under a size constraint on $\beta$ rather than to add a penalty term to the empirical score, as the former criterion makes the constraint on the coefficients more explicit.
Augmenting the empirical score by a penalty term though (as in Equation~\ref{equ:6.60}) has computational advantages, in particular to establish a criterion for choosing the penalty $\lambda$, as will be seen soon.
Whichever option is taken, it is understood from now on that, having fixed either $\lambda$ or $\mu$, the coefficients $\hat{\beta}$ depend on $\lambda$.
Clearly, assessing the suitability of a particular $\lambda$ by looking at either $R(\hat{\beta})$ or $R_{\lambda}(\hat{\beta})$ is impossible, since both measures are optimal for $\lambda = 0$.
A less optimistic measure of performance is the leave--one--out score, which is defined as follows:
Let $\beta_{\hat{\imath}}$ be the stationary point of $ \frac{1}{N-1} \sum_{k \neq i} S(g(x_k\beta^t), y_k) - \frac{\lambda}{2} |\beta|^2$, that is, we remove the $i$-th point from the training set. 
Having computed $\beta_{\hat{\imath}}$ for every $i = 1 \ldots N$, we form the leave--one--out output $g_{\hat{\imath}} := g(x_i\beta_{\hat{\imath}}^t)$ and finally the leave--one--out score
\beq{equ:6.90}
R_{\mbox{\tiny loo}}(\lambda) := \frac{1}{N} \sum_i S(g_{\hat{\imath}}, y_i),
\eeq
which is then investigated as a function of $\lambda$ (or equivalently $\mu$).
The leave--one--out score evaluates every $\beta_{\hat{\imath}}$ on precisely that  sample point which was removed from the training set before finding $\beta_{\hat{\imath}}$. 
The prospect of having to determine the coefficients $N$ times in order to compute the leave--one--out score for a single $\lambda$ seems horrible at first glance, but a few calculations will help to simplify this problem drastically.
In Appendix~\ref{apx:leave-one-out-parameters}, it will be shown that approximately
\beq{equ:6.100}
\eta_{\hat{\imath}} 
:= x_i\beta_{\hat{\imath}}^t 
\cong \hat{\eta} + \frac{1}{1 - \hat{\mtr{w}}_i x_i \mtr{H}^{-1}x_i^{t}} x_i\mtr{H}^{-1} \left[ x^t_i \hat{\mtr{d}}_i + 2 \Lambda \hat{\beta} \right].
\eeq
with 
\beq{equ:6.102}
\mtr{H} = \mtr{X}^t \hat{\mtr{W}} \mtr{X} + (N-1) \Lambda, 
\qquad
\Lambda = 
\left( \begin{array}{cccc}
 0		& 			& 			& 0			\\
			& 1 		& 			& 0			\\
			& 			& \ddots&				\\
 0 		& 0			&				& 1			
\end{array}
\right)
\eeq
All quantities that carry a hat $\hat{}$ are evaluated at $\hat{\beta}$.
The right hand side of Equation~\eqref{equ:6.100} is a function of $\hat{\beta}$ and $\lambda$ and therefore can be calculated {\em without} having to determine all $N$ leave--one--out coefficients explicitely.
Furthermore, to compute $\eta_{\hat{\imath}}$, only very few operations are required repeatedly for every $i$. 
The matrix inversion $\mtr{H}^{-1}$ needs to be performed only once.
In fact, if the Newton--Raphson method is used, all quantities can be recycled.
The leave--one--out error will in general be larger than $R(\hat{\beta})$.
The difference allows for a very interesting interpretation in terms of effective degrees of freedom of the model. 
We now proceed assuming that the Ignorance has been used as a score.
Define $\delta$ by
\beq{equ:6.105}
R_{\mbox{\tiny loo}}(\lambda) = R(\hat{\beta}) + \frac{\delta}{N},
\eeq
It is possible to show~\citep{stone77} that for a model with free parameters, $\delta$ asymptotically equals the dimension of the parameter space.
Using similar calculations in the present case, it is possible to show that $\delta \cong d + 1 - O(\lambda) $ for small $\lambda$ and $\delta \to 1 $ for large $\lambda$, where $d$ is the number of parameters in the model (not counting the intercept).
If we interprete $\delta$ as the number of effective degrees of freedom of the model, we obtain the reassuring conclusion that for vanishing penalty the model has $d+1$ effective degrees of freedom, while for increasing penalty the number of effective degrees of freedom reduces to~$1$, owing to the fact that no penalty was imposed on the intercept $\beta_0$.
Equation~\eqref{equ:6.105} is a version of Akaike's information criterion ~\citep[AIC, see e.g.][]{hastie01}.
Akaike recommends that if models are indexed by a parameter $\lambda$, say, the model with minimum
\beq{equ:6.190}
\mbox{AIC} := 2 R(\hat{\beta}) + 2 \* \frac{\delta}{N}.
\eeq
should be selected, which we see is asymptotically equivalent to minimising $R_{\mbox{\tiny loo}}$.
Although AIC and leave--one--out error are asymptotically the same, the two quantities can differ somewhat for very small sample sizes. 
If the degrees of freedom of a model are known for some reason, it is possible to use the AIC directly as a criterion for determining the regularisation parameter.
We will however go the other way and, knowing $R_{\mbox{\tiny loo}}$ and $R(\hat{\beta})$, determine $\delta$ for diagnostic purposes. 
There is a corresponding relation for the Brier score, relating $R_{\mbox{\tiny loo}}$, $R(\hat{\beta})$ and the degrees of freedom.
This statistic, known as $C_p$~statistic\cite{hastie01}, is given by
\beq{equ:6.200}
C_p := R(\hat{\beta}) + 2\frac{\delta}{N} \* \Ex [g(1-g)] .
\eeq
The derivation of $C_p$ assumes that $g(1-g)$ is approximately constant. 
This might be justified in many {\em linear} regression situations or if $g$ has a sharply concentrated distribution, but in general this seems to be a quite idealistic assumption.
A direct calculation though of the leave--one--out error Equation~\eqref{equ:6.90} with the approximation~\eqref{equ:6.100}, both valid for any score, does not suffer from these problems and were found here to give much better results.
As said, the AIC or the $C_p$--statistic might still be a last resort if calculating the leave--one--out coefficients $\beta_{\hat{\imath}}$ is difficult or impossible. 
It is then necessary to obtain $\delta$, the number of effective degrees of freedom, by other means. 
This can require tricky analysis.
The next subsection discusses an interesting modification of the current setup for which the number of degrees of freedom are fortunately known.
\subsection{$L_1$--Type Regularisation, or the Lasso}
\label{subsec:l_1-type-regularisation}
In the previous subsection, we regularised our estimates by constraining the size of $\beta$, measured in the $L_2$--sense. 
For linear regression, \citet{tibshirani96} suggested to use the $L_1$--norm as an alternative, that is, the score is minimised under the constraint $|\beta| \leq \mu$, where $|\beta| := \sum_{k = 1}^{d} |\beta_k|$.
As before, no constraint is placed on the intercept.
The resulting regression technique has become known as the lasso.
An interesting feature of the lasso though is that with increasingly tight constraining, some coefficients become {\em exactly} zero. 
Interpreting the corresponding inputs as ``less important'', the lasso technique is appealing also from a diagnostic point of view.
Recently, it has been shown by~\cite{zou07} that, consistent with intuition, the number of degrees of freedom of the lasso is given by the number of nonzero coefficients.
The main features of the lasso persist when logistic models are used with an $L_1$--penalty on the coefficients, that is, with increasingly tight constraining, some of the coefficients vanish exactly.
The name ``lasso'' will be kept also for the logistic case, even though it was originally used for the linear case only.
Calculating the coefficients for the lasso is more involved than for standard regression or $L_2$--regularised regression and requires quadratic optimisation techniques.
We have not rigorously proven that the number of degrees of freedom in the logistic case is still given by the number of nonzero coefficients, although this appears to be quite plausible.
Hence we suggest to determine the regularisation parameter by minimising 
\beq{equ:6.210}
\mbox{AIC} = 2 R(\hat{\beta}) + 2 \* \frac{\delta}{N},
\eeq
where $\delta$ is the number of nonzero coefficients, $R$ is the empirical Ignorance score, and $\hat{\beta}$ is the coefficient which optimises $R(\beta)$ under the constraint $|\beta| \leq \mu$ or equivalently $R(\beta) + \lambda |\beta|$.
\section{Numerical Experiments}
\label{sec:numerical_studies}
\begin{figure}
\begin{center}
\begin{tabular}{r}
\epsfig{file = 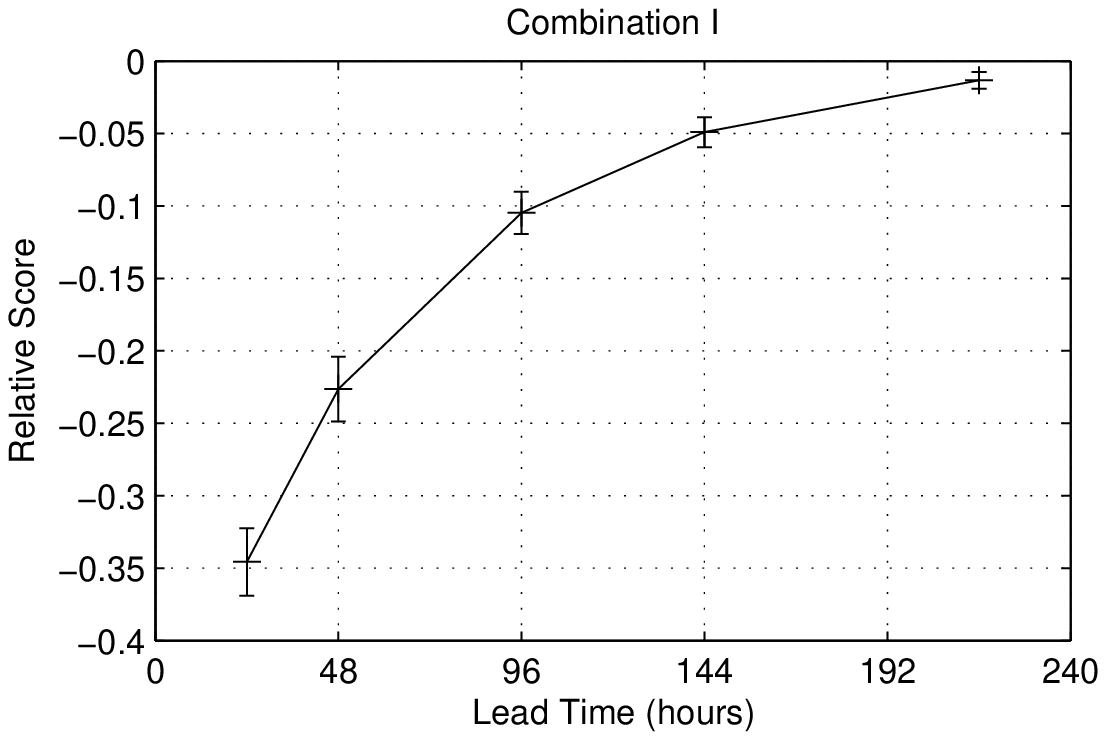}\\
\epsfig{file = 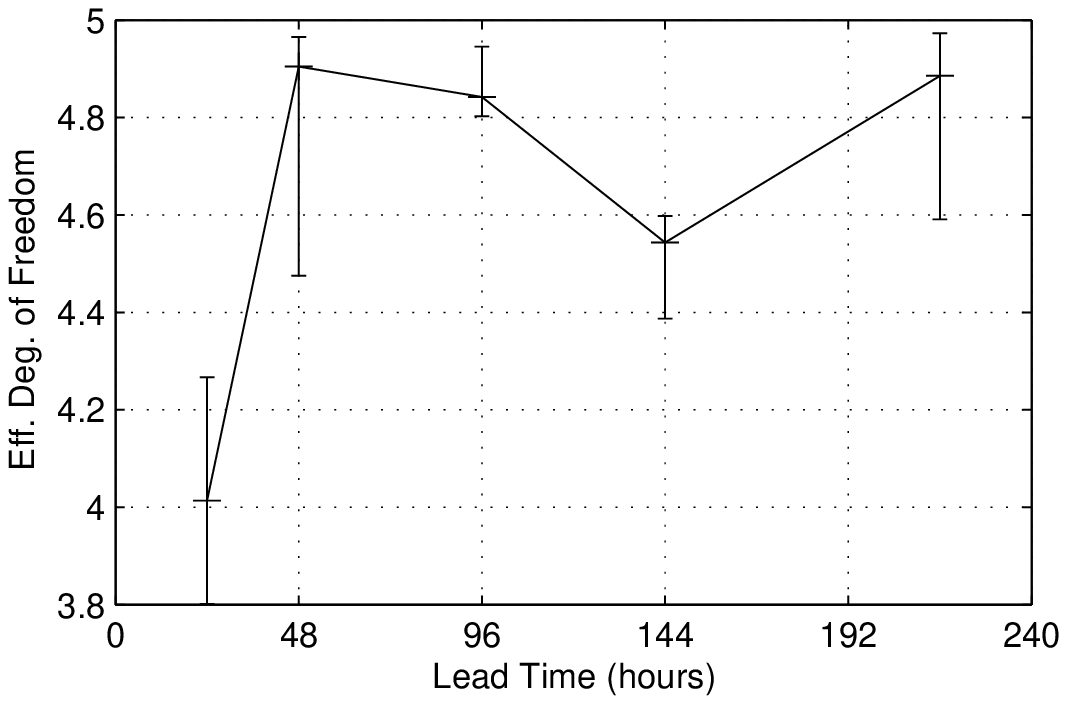}
\end{tabular}
\end{center}
\caption{\label{fig:logit-skill-stwmo10015feat-set1}%
Empirical Ignorance score (top panel) and effective degrees of freedom~$\delta$ (bottom panel) for input combination~I (see table~\ref{tab:feature-sets}). 
The reference here is climatology.
The confidence bars were obtained using tenfold cross validation.}
\end{figure}

\begin{figure}
\begin{center}
\begin{tabular}{r}
\epsfig{file = 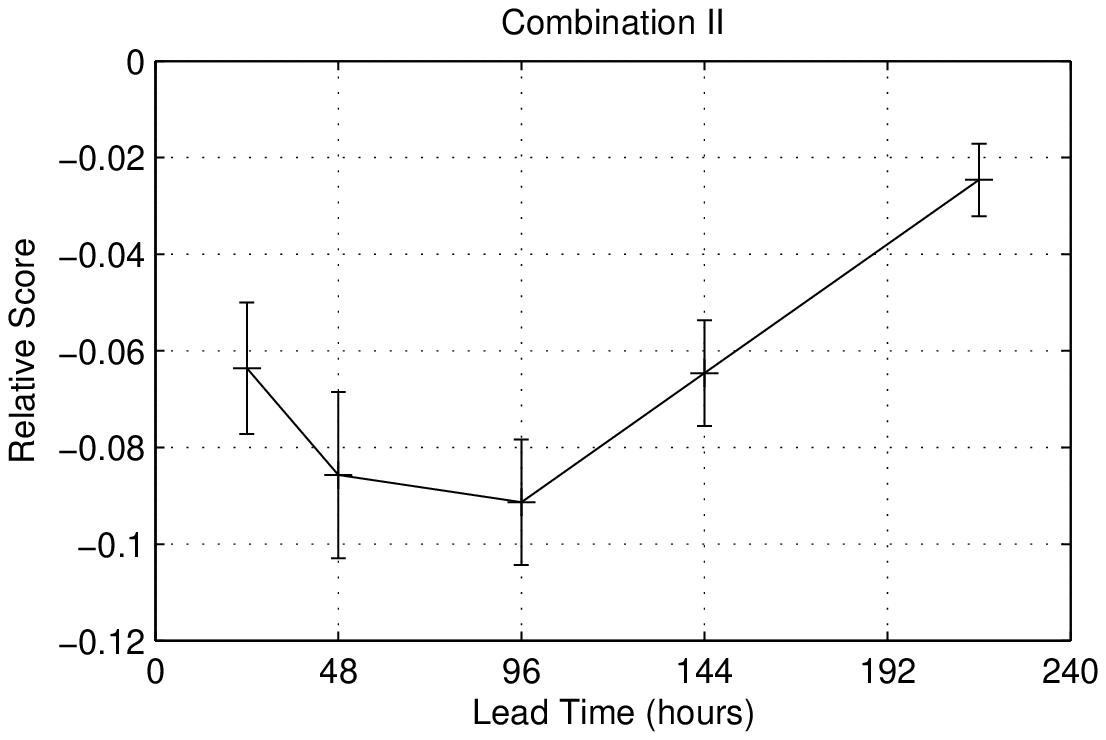}\\
\epsfig{file = 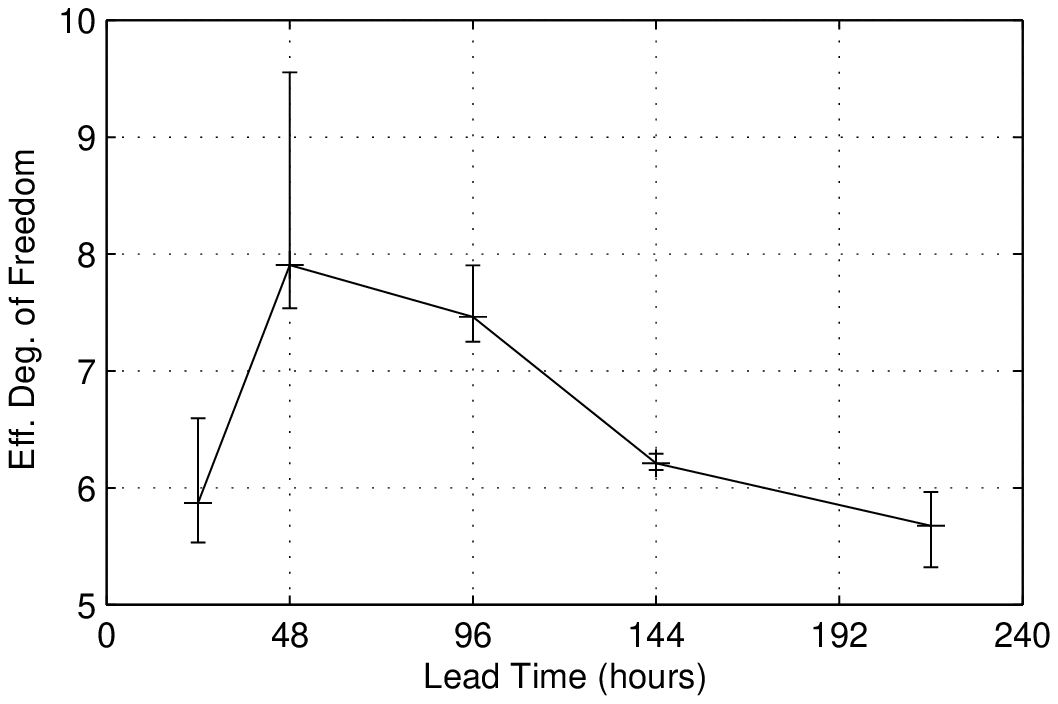}
\end{tabular}
\end{center}
\caption{\label{fig:logit-skill-stwmo10015feat-set3}%
Empirical Ignorance score (top panel) and effective degrees of freedom~$\delta$ (bottom panel) for input combination~II (see table~\ref{tab:feature-sets}). 
The score is presented with combination~I as a reference.
The confidence bars were obtained using tenfold cross validation.}
\end{figure}

\begin{figure}
\begin{center}
\begin{tabular}{r}
\epsfig{file = 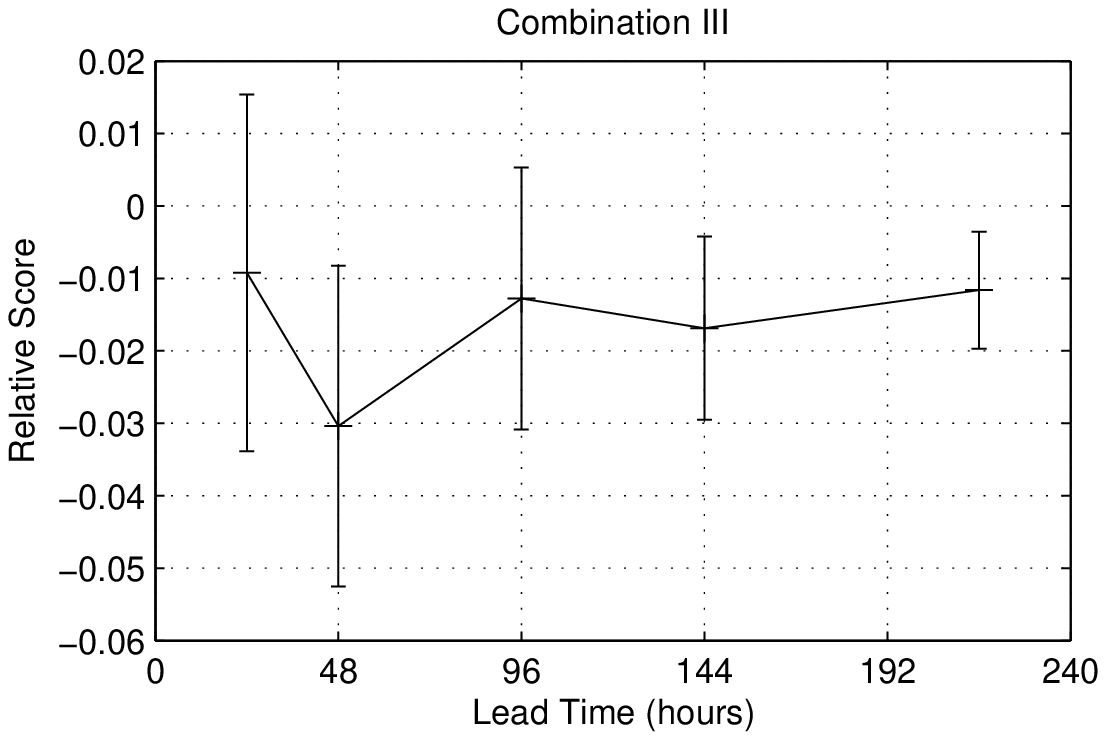}\\
\epsfig{file = 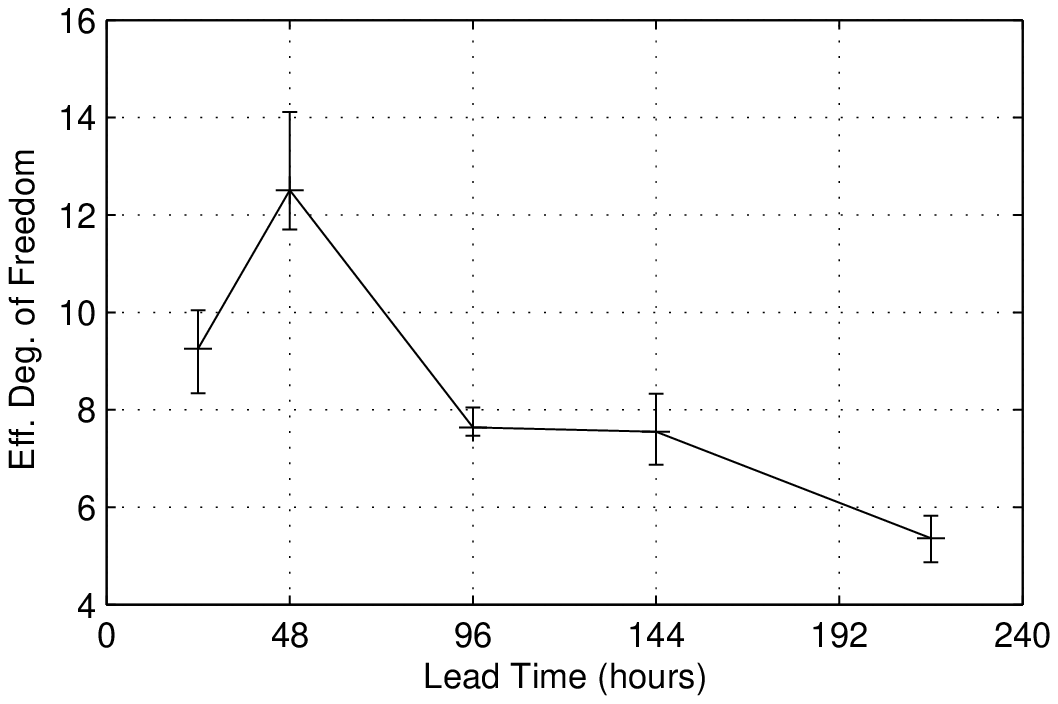}
\end{tabular}
\end{center}
\caption{\label{fig:logit-skill-stwmo10015feat-set2}%
Empirical Ignorance score (top panel) and effective degrees of freedom~$\delta$ (bottom panel) for input combination~III (see table~\ref{tab:feature-sets}). 
The score is presented with combination~II as a reference.
The confidence bars were obtained using tenfold cross validation.}
\end{figure}

\begin{figure}
\begin{center}
\begin{tabular}{r}
\epsfig{file = 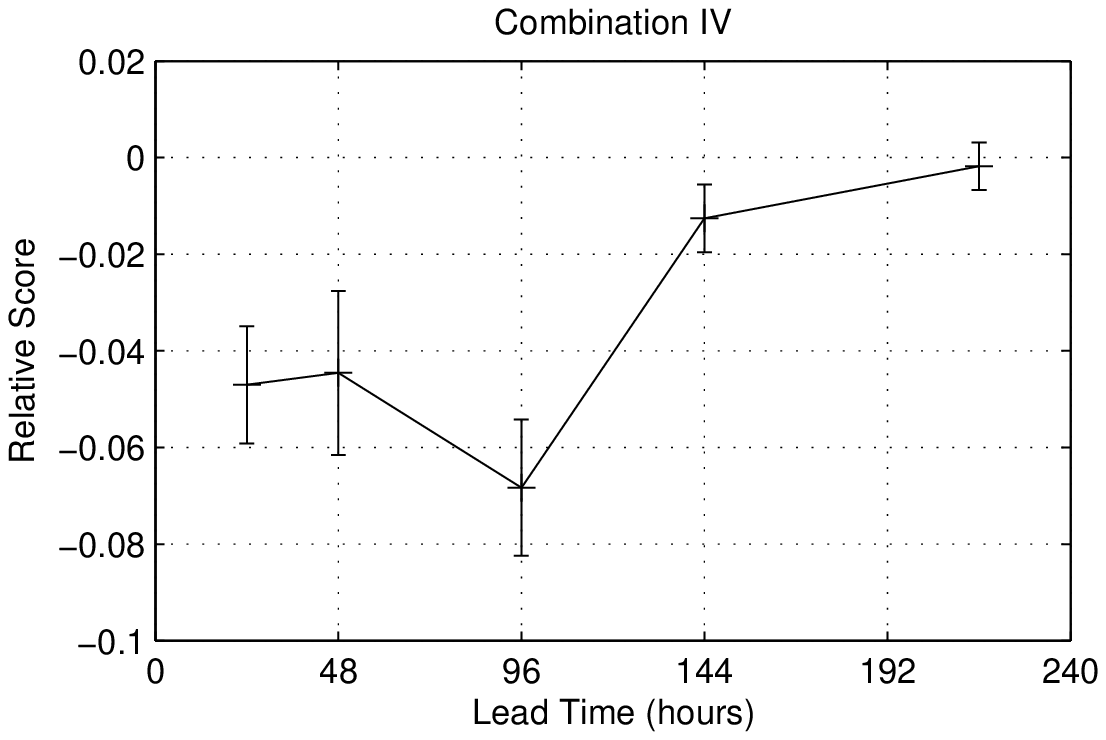}\\
\epsfig{file = 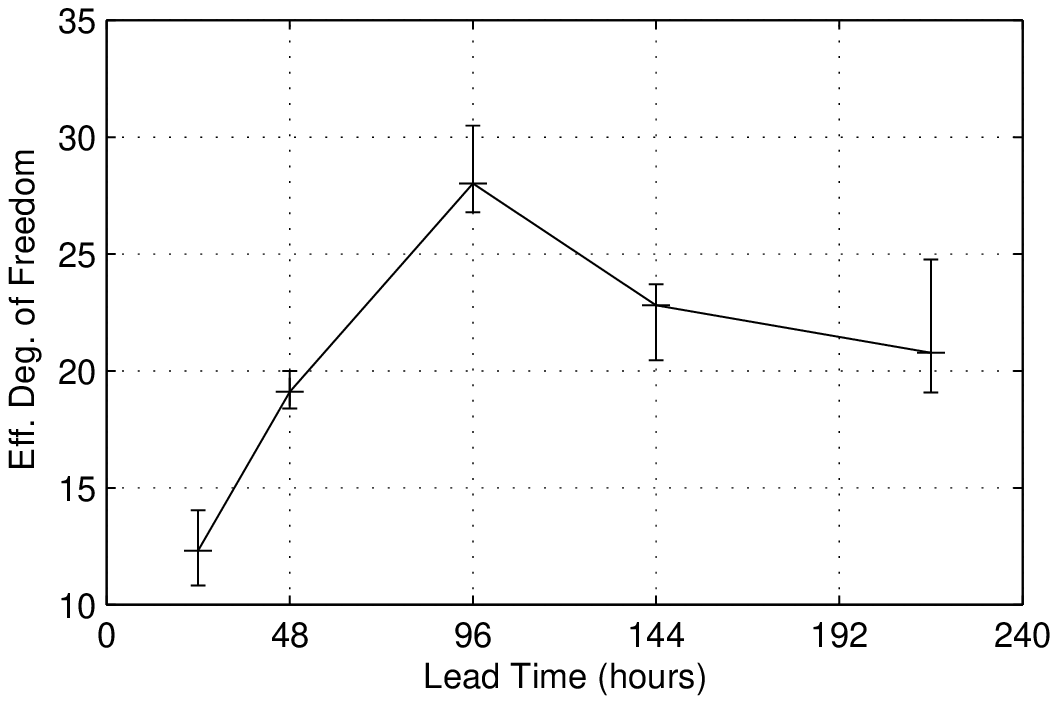}
\end{tabular}
\end{center}
\caption{\label{fig:logit-skill-stwmo10015feat-set4}%
Empirical Ignorance score (top panel) and effective degrees of freedom~$\delta$ (bottom panel) for input combination~IV (see table~\ref{tab:feature-sets}). 
The score is presented with combination~III as a reference.
The confidence bars were obtained using tenfold cross validation.}
\end{figure}

\begin{figure}
\begin{center}
\epsfig{file = 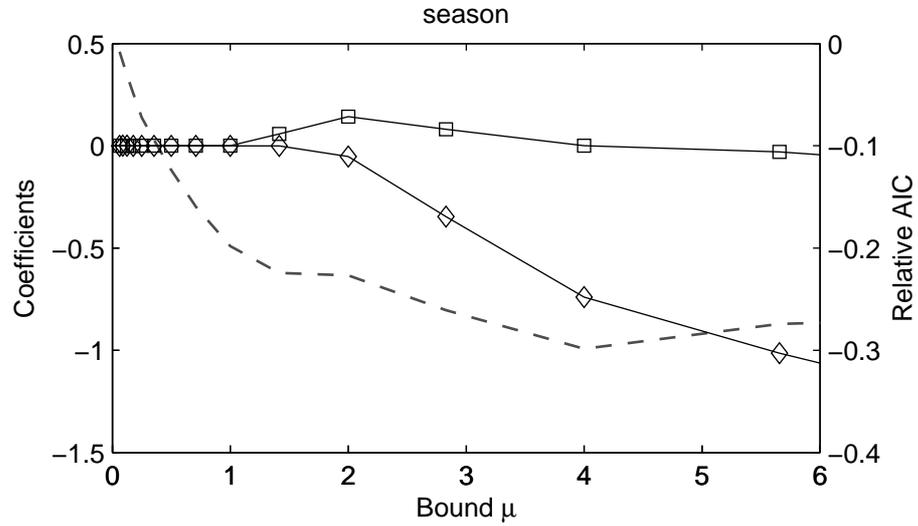}
\end{center}
\caption{\label{fig:lasso-lt48stwmo10015varseason}%
The coefficients corresponding to the input \texttt{season} for a lasso model, plot over the regularisation parameter $\mu$.
Here, $\Box = \cos(\ldots)$ and $\Diamond = \sin(\ldots)$ of Equation~\ref{equ:7.10}.}
\end{figure}

\begin{figure}
\begin{center}
\epsfig{file = 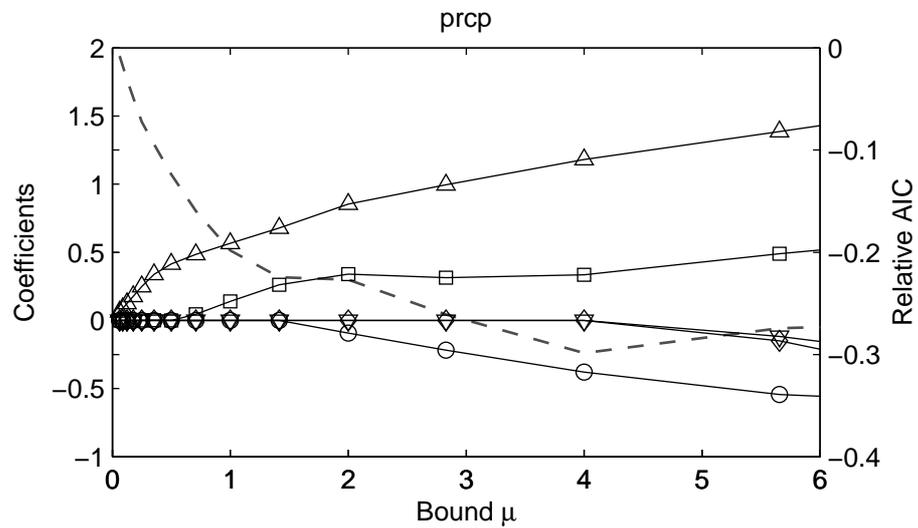}
\end{center}
\caption{\label{fig:lasso-lt48stwmo10015varprcp}%
The coefficients corresponding to the inputs related to \texttt{prcp} (precipitation) for a lasso model, plot over the regularisation parameter $\mu$.
Here, $\Box =$~High Resolution Forecast, $\Diamond =$~(High Resolution Fc.)$^2$, $\bigtriangleup =$~Ensemble Mean, $\bigtriangledown =$~(Ensemble Mean)$^2$, $\bigcirc =$~Ensemble variance.}
\end{figure}

\begin{figure}
\begin{center}
\epsfig{file = 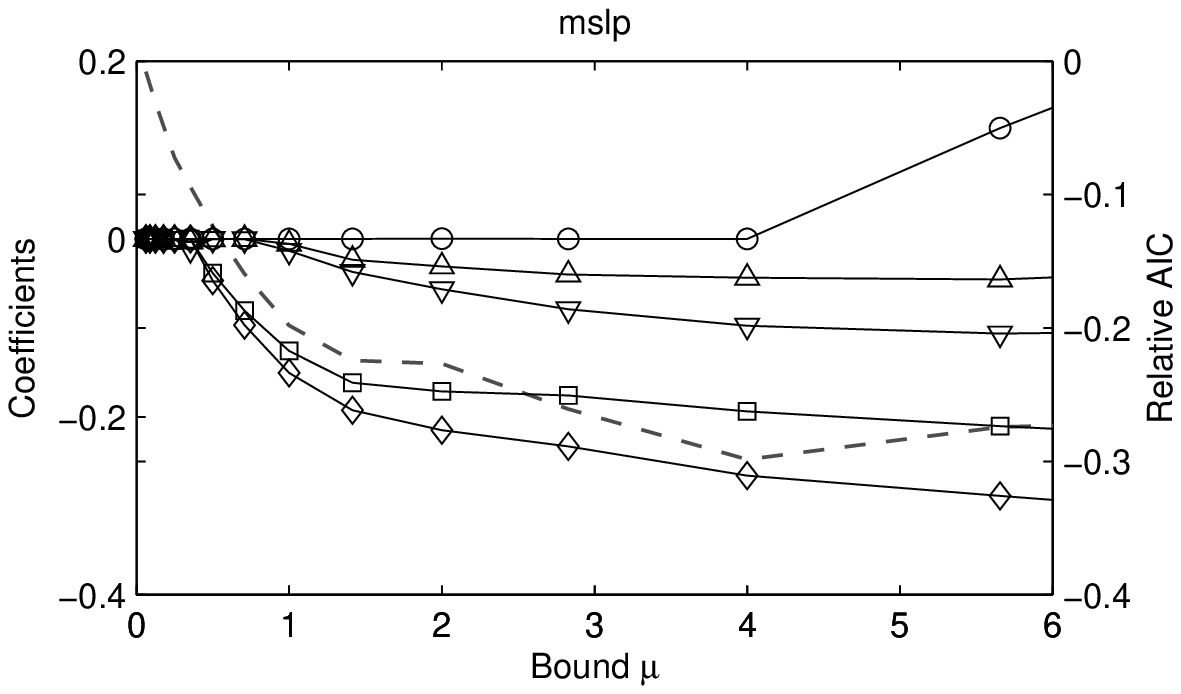}
\end{center}
\caption{\label{fig:lasso-lt48stwmo10015varmslp}%
The coefficients corresponding to the inputs related to \texttt{mslp} (mean sea level pressure) for a lasso model, plot over the regularisation parameter $\mu$.
For explanation of line markers see Figure~\ref{fig:lasso-lt48stwmo10015varprcp}.}
\end{figure}

\begin{figure}
\begin{center}
\epsfig{file = 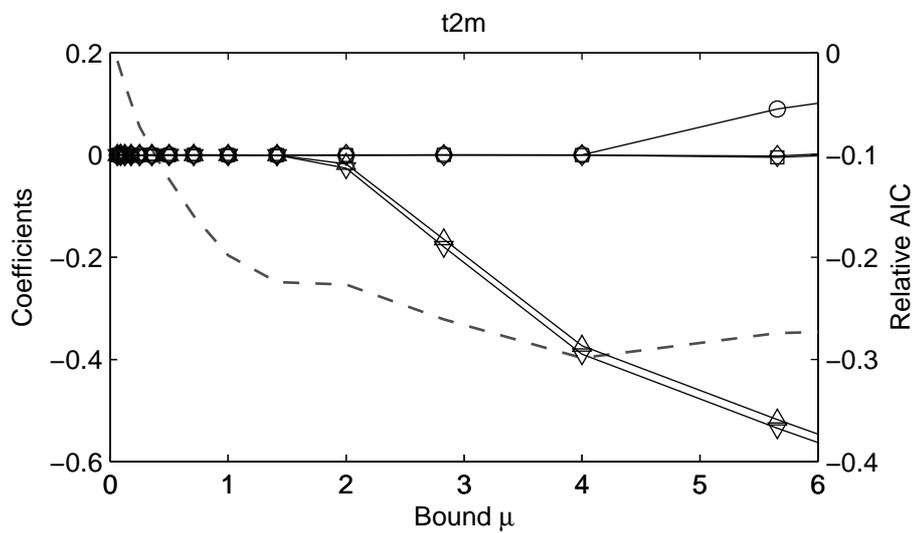}
\end{center}
\caption{\label{fig:lasso-lt48stwmo10015vart2m}%
The coefficients corresponding to the inputs related to \texttt{t2m} (temperature) for a lasso model, plot over the regularisation parameter $\mu$.
For explanation of line markers see Figure~\ref{fig:lasso-lt48stwmo10015varprcp}.}
\end{figure}

\begin{figure}
\begin{center}
\epsfig{file = 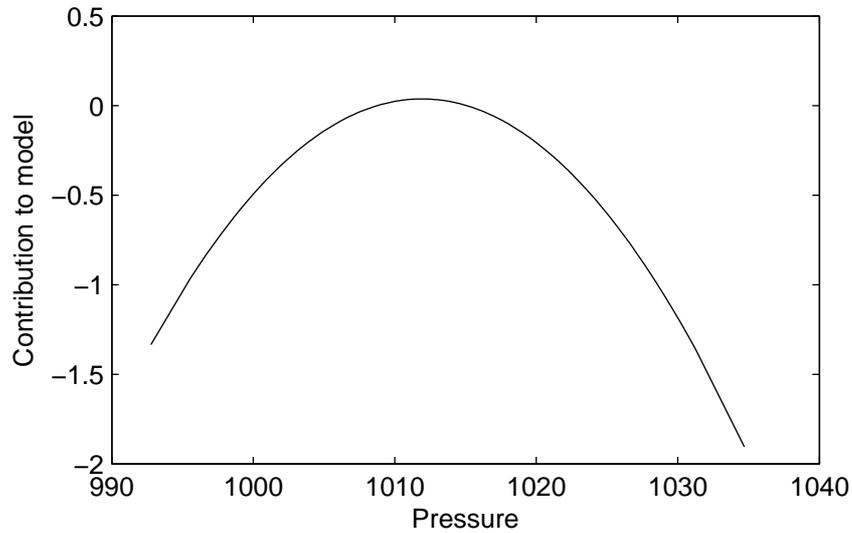}
\end{center}
\caption{\label{fig:pressure}Contribution of the \texttt{mslp} (pressure) high resolution forecast to the lasso model, more specifically, to the logarithmic odds ratio. 
There is a significant contribution from the term quadratic in \texttt{mslp}.}
\end{figure}

In this section, regularised logistic regression is applied to the occurrence of precipitation.
Several weather stations in Western Europe were investigated, with similar findings. 
As a representative example, results for Heligoland,~Germany~(WMO~10015) are presented here. 
As inputs, high resolution deterministic and ensemble forecasts were used.
The ensemble forecasts consist of the 50 (perturbed) member ensemble, produced by the then--operational ECMWF global ensemble prediction system.
Station data of precipitation was kindly provided by ECMWF as well.
Forecasts were available for the years 2001--2005, featuring lead times from one to ten days.
All data verified at noon.
To form the inputs, high resolution, control and ensemble forecasts for mean~sea~level~pressure, two~metre~temperature, and precipitation itself were used.
Different combinations of inputs were tested.
To describe the input combinations, we will use the following abbreviations:
We write \texttt{prcp}, \texttt{mslp}, and \texttt{t2m} for precipitation,  mean~sea~level~pressure, and two~metre~temperature, respectively.
The high resolution forecast is indicated with a suffix~\texttt{\_h}, while the control and the ensemble carry the suffixes~\texttt{\_c} and~\texttt{\_e}, respectively.
For example, the high resolution mean sea level pressure forecast is denoted by \texttt{mslp\_h}, or the ensemble two metre temperature forecast by \texttt{t2m\_e}.
The input \texttt{season} is simply the phase of the year, that is, 
\beq{equ:7.10}
\mathtt{season} = \left( \cos(\frac{2 \pi}{365.2425} n), \sin(\frac{2 \pi}{365.2425} n) \right) \qquad n = \mbox{no.\ of the day}
\eeq
In total, four different combinations were investigated (see Table~\ref{tab:feature-sets}).
All combinations include \texttt{season}.
The first combination adds \texttt{prcp\_h}, resulting in 3~inputs.
The second combination uses all available high resolution forecasts twice: plain and squared (8~inputs), thereby modelling potential nonlinear connections between precipitation events and the inputs.
The third  combination uses all available precipitation forecasts: \texttt{prcp\_h}, \texttt{prcp\_c}, and \texttt{prcp\_e} (54~inputs).
The fourth combination uses all available forecasts: high resolution, control and ensemble for precip, pressure and temperature.
Again, each forecast is included both plain and squared.
This combination comprises 314~inputs. 
I should say that, to the best of my knowledge (although I cannot claim to have combed the literature very thoroughly), so far only the ensemble mean and spread have been used as inputs to logistic regression.
This might have been done either to avoid over-fitting or in the belief that the ensemble does not contain relevant information beyond mean and spread.
The results for the four combinations are displayed in Figures~\ref{fig:logit-skill-stwmo10015feat-set1}--\ref{fig:logit-skill-stwmo10015feat-set4}, which show the empirical score (top panels) and the number of effective degrees of freedom (bottom panels), as defined in Equation~\eqref{equ:6.105}.
The performance is presented in an incremental fashion:  Figure~\ref{fig:logit-skill-stwmo10015feat-set1}, top panel, shows the performance of combination~I relative to climatology, Figure~\ref{fig:logit-skill-stwmo10015feat-set3}, top panel, shows the performance of combination~II relative to combination~I and so forth.
The confidence bars for all plots were obtained using 10--fold cross validation.
Figure~\ref{fig:logit-skill-stwmo10015feat-set1}, top panel, demonstrates the interesting (albeit maybe not surprising) fact that the high resolution forecast for precipitation contains a fair amount of probabilistic information, if processed correctly.
The forecast even seems to have skill out to day~9.
Since this model is trained on around 1600 samples but has only four parameters, it is not surprising that the number of effective degrees of freedom $\delta$ is between 4~and~5.
The somewhat odd finding that $\delta$ is even larger than the total number of parameters has two reasons.
Firstly, we are dealing with a finite data set, while the result is true only in the limit of an infinite amount of data.
Secondly, due to the penalisation, our parameters are not asymptotically unbiased. 
Strictly speaking, this entails a further correction to Equation~\eqref{equ:6.105}, which we found to be less than .1 in all considered examples.
Adding high resolution forecasts for mean sea level pressure and temperature to the mix generally adds skill, as Figure~\ref{fig:logit-skill-stwmo10015feat-set3}, top panel, demonstrates.
There seems to be no effect though at high lead times, like 9~days.
We will see later that there is strong indication that the squared mean sea level pressure adds additional information, more specifically the high resolution \texttt{mslp\_h}$^2$ at short lead times and the ensemble \texttt{mslp\_e}$^2$ at longer lead times.
What we can see already from Figure~\ref{fig:logit-skill-stwmo10015feat-set3}, bottom panel, though is that the number of EDF's increases, in particular for lead times 48h and 96h, where this combination features the largest increase in skill over combination~I.
The number of effective degrees of freedom $\delta$ is however always significantly smaller than~9, the number of coefficients in this model. 
This indicates that not all of the additional inputs add independent information.
Somewhat surprisingly, the precipitation ensemble (along with \texttt{prcp\_h} and \texttt{prcp\_c}) shows close to no improvement in skill over combination~I in this context, apart from lead time~48h maybe (Fig.~\ref{fig:logit-skill-stwmo10015feat-set2}, top panel).
The number of effective degrees of freedom $\delta$ is always far less than the number of coefficients in this model, and from lead times 96h~onwards, $\delta$ is even comparable to what was found for combination~II.
Significant increase in skill is obtained using combination~VI (Fig.~\ref{fig:logit-skill-stwmo10015feat-set4}, top panel).
In addition to combination~III, the present combination uses pressure and temperature as well as all variables once plain and once squared.
It would of course be interesting to know which of the inputs makes the biggest difference. 
We have not investigated this in full, but a partial answer will be obtained later using the lasso technique.
What is important here is that despite the large number of coefficients~(314), there appear to be no signs of over-fitting. 
Figure~\ref{fig:logit-skill-stwmo10015feat-set4}, bottom panel, demonstrates that the number of effective degrees of freedom $\delta$, albeit always far less than the number of coefficients in this model, is significantly larger than for any other combinations tested here.
We can conclude that the additional inputs in fact do add additional information.
An interesting aside is that $\delta$ always seems to have a maximum around 96h~lead time. 
We have not investigated the reason, but speculate that this is related to the ensemble generation.
The ensembles are free runs of slightly perturbed initial conditions. 
The spread of the ensemble typically grows initially, due to the local instabilities. 
The growth of uncertainty is of course the reason why ensembles are used in the first place. 
Eventually, the spread will saturate due to nonlinear effects. The middle ground is where each ensemble member adds the most information to the whole.
We are now going to discuss some sample results for the lasso.
The goal is to get some idea as to the relative importance of the inputs.
As discussed, an interesting feature of the lasso is that with increasingly tight bound on the coefficients (i.e.~decreasing $\mu$), some coefficients become zero.
The numerical experiments discussed here readily demonstrate this effect.
As an example, forecasts for lead time~48h were considered.
As inputs the variables \texttt{prcp}, \texttt{mslp}, \texttt{t2m}, were used, along with \texttt{season}.
To get the big picture, we dispensed with including all available types of forecasts but concentrated on the high resolution forecasts, the high resolution forecast squared, the ensemble mean, the ensemble mean squared, and the ensemble variance\footnote{Alternatively, one could take the the mean of the squared ensemble members, but choosing the variance makes the model's dependence on the ensemble spread explicit.}.
In fact, there was enough data available so that in this case, the coefficients could have been determined without regularisation.
In order to see regularisation ``in action'', we reduced the data set to only 6 months. 
The coefficients as a function of the bound $\mu$ are displayed in Figures~\ref{fig:lasso-lt48stwmo10015varseason}--\ref{fig:lasso-lt48stwmo10015vart2m}.
In order to avoid clutter, the coefficients corresponding to different variables are displayed in different figures.
All plots show the AIC for reference (dashed grey line and right ordinate).
The AIC shows a minimum at around $\mu = 4$, so this would be the bound chosen by the AIC criterion.
At this point, there are 10~nonzero coefficients.
In terms of importance of individual inputs, the magnitude of the coefficients is probably not as descriptive as the point at which a given coefficient drops out of the mix with increasingly tight bound.
The longest--surviving inputs are (in descending order) the \texttt{prcp} ensemble mean, the \texttt{mslp} high resolution squared and the \texttt{mslp} high resolution.
The \texttt{prcp} high resolution seems to be almost as important.
\texttt{t2m} gets coefficients of comparable magnitude, but drops out earlier.
The behaviour of \texttt{prcp} ensemble variance is similar.
The most interesting observation is probably the contribution of the \texttt{mslp} high resolution forecast to the model, displayed in Figure~\ref{fig:pressure}.
Reassuringly, we obtain that the probability of rain decreases with both increasing and decreasing pressure, with a maximum at around 1012hPa.
Of course, this analysis neglects the influence of the other inputs, which are certainly not independent of pressure.
In addition, Figure~\ref{fig:lasso-lt48stwmo10015varmslp} shows that the contribution of the \texttt{mslp} ensemble mean is very similar to that of the \texttt{mslp} high resolution forecast, albeit with smaller magnitude.
It turns out that the importance of the \texttt{mslp} ensemble mean increases with increasing lead time (not shown), with high resolution and ensemble mean effectively reversing role at around lead time~120h.
%
%
%
%
%
%
%
%
%
%
%
%
%
%
%
%
%
%
\begin{table}
\begin{center}
\begin{tabular}{lll}
 & Inputs & No.~of Coeffz.\\[1mm] 
I & \texttt{season, prcp\_h} & 3 \\[1mm] 
II & \texttt{season, prcp\_h, prcp\_h$^2$, mslp\_h, mslp\_h$^2$}, & \\
 &  \texttt{t2m\_h, t2m\_h$^2$} & 8 \\[1mm] 
III & \texttt{season, prcp\_h, prcp\_c, prcp\_e} & 54 \\[1mm] 
IV & \texttt{season, $*$\_h, $*$\_c, $*$\_e, $*$\_h$^2$, $*$\_c$^2$, $*$\_e$^2$} & \\
 & with $*=$ \texttt{prcp, mslp, t2m} & 314
\end{tabular}
\end{center}
\caption{\label{tab:feature-sets} Tested combination of inputs. The number of coefficients does not include the intercept.}
\end{table}

\section{Conclusion}
\label{sec:conclusion}
Logistic models were considered as a statistical tool to map both deterministic and ensemble weather forecasts into probabilities. 
The parameters of the logistic model were found by optimising proper scoring rules.
Due to the variety of different forecast products that are typically available for a single target, the number of potential inputs to the logistic model might be large, comparable even to the number of forecast instances.
This number is increased even further if, next to the original inputs themselves, some nonlinear functions of them are to be included (for example squared terms or bilinear combinations of two inputs). 
This renders a reliable determination of the model coefficients difficult due to over-fitting.
This paper presents systematical approaches to avoid over-fitting in logistic models.
Motivated by similar approaches for linear regression, the coefficients are penalised, thereby reducing their potential variance. 
Techniques for model selection using efficient leave--one--out cross validation are discussed. 
As an example, probabilities of precipitation are considered. 
Not only forecasts for precipitation, but also pressure and temperature are used (both as high resolution as well as ensemble forecasts).
It is demonstrated that these forecasts do add information to the pure precipitation forecasts.
Despite the large number of inputs, it is demonstrated that the parameters can be determined in a reliable fashion.
A variant of the logistic model called the lasso was discussed. 
In this model, constraining the coefficients can result in some of them being exactly zero, allowing to identify them as less important. 
A notable observation here is the large contribution of the \texttt{mslp} high resolution forecast itself and its square for small lead times, and similarly of the \texttt{mslp} ensemble forecast for larger lead times.

\section*{Acknowledgements}
Fruitful discussions with Leonard A.~Smith (London School of Economics) as well as Ivan G.~Szendro Ter\'{a}n and Holger Kantz (Max--Planck--Institut f\"{u}r Physik komplexer Systeme, Dresden) are acknowledged.
I am also indebted to Christian Merkwirth (UJ~Krak\'{o}w) and J\"{o}rg Wichard (FMP--Berlin) for revealing to me various secrets of statistical learning.
Finally, I would like to thank the European Centre For Medium Range Weather
Forecasts (ECMWF), in particular Renate Hagedorn, for kindfully providing forecasts and verification data.
\appendix
\section{Leave--one--out parameters}
\label{apx:leave-one-out-parameters}
In this appendix, I want to demonstrate that Equation~\ref{equ:6.100} is a good approximation for the linear response to the input $x_i$ of the model with  parameters $\beta_{\hat{\imath}}$, that is, which are based on a training set without the $i$--th input--target pair.
I will use the following abbreviations (consistent with the notation elsewhere in the paper):
\begin{eqnarray}
\label{equ:apx.10}
\eta_i & := & x_i\beta^t \qquad \mbox{for all } i = 1 \ldots N \\
\label{equ:apx.20}
g_i & := & g(\eta_i) \qquad \mbox{for all } i = 1 \ldots N \\
\label{equ:apx.30}
R(\beta) & := & \frac{1}{N} \sum_{k = 1}^N S(g_k, y_k)\\
\label{equ:apx.40}
R_{\lambda}(\beta) & := & R(\beta) + \lambda |\beta|^2,
\end{eqnarray}
The reader be again reminded that the penalty term $|\beta|^2 = \sum_{k = 1}^d \beta_k^2$ does not involve the intercept $\beta_0$.
Let $\hat{\beta}$ be a stationary point of $R_{\lambda}(\beta)$, that is $\partial R_{\lambda}(\hat{\beta}) / \partial \beta = 0$.
In analogy to the definitions~\eqref{equ:apx.10} and~\eqref{equ:apx.20}, I set $\hat{\eta_i} := x_i\hat{\beta}^t$, $\hat{g_i} := g(\hat{\eta_i})$ for all $i = 1 \ldots N$.
Let $\beta_{\hat{\imath}}$ be a stationary point of $R_{\lambda}(\beta)$ but omitting the $i$--th input--target pair, or more specifically, 
\beq{equ:apx.50}
0 = \partial_{\beta} \left[ \frac{1}{N-1} \sum_{k \neq i} S(g_k, y_k) + \lambda |\beta|^2 \right]_{\beta = \beta_{\hat{\imath}}}.
\eeq
Now from the definition of $R_{\lambda}$ we get
\beq{equ:apx.60}
N R_{\lambda}(\beta) 
= \sum_{k = 1}^N S(g_k, y_k) + N \lambda |\beta|^2
= \sum_{k \neq i} S(g_k, y_k) + (N - 1) \lambda |\beta|^2 + \lambda |\beta|^2 + S(g_i, y_i).
\eeq
We now take the derivative of this at $\beta_{\hat{\imath}}$ and divide by $N$. 
Using Equation~\eqref{equ:apx.50} we obtain
\beq{equ:apx.70}
\partial_{\beta} R_{\lambda}(\beta_{\hat{\imath}}) 
= \frac{1}{N} \partial_{\beta}\left[  S(g_i, y_i) 
	+ \lambda |\beta|^2 \right]_{\beta = \beta_{\hat{\imath}}}.
\eeq
On the other hand, we can expand $ \partial_{\beta} R_{\lambda} (\beta)$ to first order in $\beta_{\hat{\imath}} - \hat{\beta}$ at $\hat{\beta}$
\beq{equ:apx.80}
\partial_{\beta} R_{\lambda}(\beta_{\hat{\imath}}) 
\cong \underbrace{\partial_{\beta} R_{\lambda}(\hat{\beta})}_{= 0}
 + \partial^2_{\beta^2} R_{\lambda}(\hat{\beta}) 
 	\left( \beta_{\hat{\imath}} - \hat{\beta}\right)^t,
\eeq
where the first term vanishes because $\hat{\beta}$ is a stationary point of $R_{\lambda}(\beta)$.  
The two relations~\eqref{equ:apx.70} and~\eqref{equ:apx.80} are central to the argumentation in this appendix.
We continue with determining $\beta_{\hat{\imath}}$.
To this end, we also linearise the first equation~\eqref{equ:apx.70}, set it equal to~\eqref{equ:apx.80}, and solve for $\beta_{\hat{\imath}} - \hat{\beta}$.
This yields
\beq{equ:apx.90}
\beta_{\hat{\imath}}^t - \hat{\beta}^t = 
\mtr{H}^{-1}_i \left[ \partial_{\beta} S(\hat{g}_i, y_i)^t + 2 \Lambda \hat{\beta}^t \right],
\eeq
with the following abbreviations:
\beq{equ:apx.100}
\Lambda := 
	\left( 
	\begin{array}{c|ccc}
		0 & & 0 & \\ \hline
		  & \lambda & & 0 \\
		0 & & \ddots & \\
		  & 0 & & \lambda
	\end{array}
	\right).
\eeq
\begin{eqnarray}
	\mtr{H}_i & = & \mtr{H} - \partial^2_{\beta^2} S(\hat{g}_i, y_i) \\
	\mtr{H} & = & N \partial^2_{\beta^2} R_{\lambda}(\hat{\beta}) - 2 \Lambda
\end{eqnarray}
We will now calculate the derivatives of $S$ in Equation~\eqref{equ:apx.90}.
Firstly, we have
\beq{equ:apx.110}
\partial_{\beta} S(\hat{g}_i, y_i) 
= \partial_{\eta} S(g(\hat{\eta}_i), y_i) \, x_i
= \hat{\mtr{d}}_i x_i
\eeq
where $\hat{\mtr{d}}_i = \partial_{\eta} S(g(\hat{\eta}_i), y_i)$ is used as a shorthand (consistent with Equation~\ref{equ:6.40}).
Secondly, we have
\beq{equ:apx.120}
\partial^2_{\beta^2} S(\hat{g}_i, y_i) 
= \hat{\mtr{w}}_i x^t_i x_i
\eeq
where $\hat{\mtr{w}}_i = \partial^2_{\eta^2} S(g(\hat{\eta}_i), y_i)$ is used as a shorthand (consistent with Equation~\ref{equ:6.45}).
With $\hat{\mtr{W}}_{kl} = \delta_{kl} \hat{\mtr{w}}_k$ and $\mtr{X}_{kl} = x^{(l)}_k$ we get
\beq{equ:apx.130}
\mtr{H} = \mtr{X}^t\hat{\mtr{W}}\mtr{X} + 2(N-1)\Lambda, \qquad
\mtr{H}_i = \mtr{H} - \hat{\mtr{w}}_i x^t_i x_i.
\eeq
It is not necessary to invert the full matrix $\mtr{H}_i$ for every $i$.
Once the inverse of $\mtr{H}$ is known, applying the Sherman--Morrison formula (see e.g.~\cite{golub96}) gives
\beq{equ:apx.140}
\mtr{H}_i^{-1} = \mtr{H}^{-1} + \frac{\hat{\mtr{w}}_i }{1 - \hat{\mtr{w}}_i x_i \mtr{H}^{-1}x_i^{t}} \mtr{H}^{-1} x_i^{t} x_i \mtr{H}^{-1}.
\eeq
Consequently, 
\beq{equ:apx.150}
x_i\mtr{H}_i^{-1} = \frac{1}{1 - \hat{\mtr{w}}_i x_i \mtr{H}^{-1}x_i^{t}}x_i\mtr{H}^{-1}.
\eeq
Since $\eta_{\hat{\imath}} - \hat{\eta}_i = x_i\beta_{\hat{\imath}}^t - x_i \hat{\beta}^t$, we can employ Equation~\eqref{equ:apx.90} together with~\eqref{equ:apx.110} and~\eqref{equ:apx.150} to obtain 
\beq{equ:apx.160}
\eta_{\hat{\imath}} = \hat{\eta} + \frac{1}{1 - \hat{\mtr{w}}_i x_i \mtr{H}^{-1}x_i^{t}} x_i\mtr{H}^{-1} \left[ x^t_i \hat{\mtr{d}}_i + 2 \Lambda \hat{\beta}^t \right].
\eeq
Note that as soon as $\mtr{H}^{-1}$ has been computed, very few additional operations are required for every $i$. 
The matrix inversion $\mtr{H}^{-1}$ needs to be performed only once.
\bibliographystyle{plainnat}
\bibliography{../../../TeX/Literatur}
\end{document}